Article

# Broadband Cavity-Enhanced Absorption Spectroscopy (BBCEAS) Coupled with an Interferometer for On-Band and Off-Band Detection of Glyoxal

Callum E. Flowerday [1] , Ryan Thalman [2] , Matthew C. Asplund [1] and Jaron C. Hansen [1,*]

1  Department of Chemistry and Biochemistry, Brigham Young University, Provo, UT 84602, USA
2  Department of Chemistry, Snow College, Richfield, UT 84701, USA; ryan.thalman@snow.edu
*  Correspondence: jhansen@chem.byu.edu; Tel.: +1-801-422-4066

**Abstract:** Glyoxal (CHOCHO) is a trace gas in the atmosphere, often used as an indicator of biogenic emissions. It is frequently compared to formaldehyde concentrations, which serve as indicators of anthropogenic emissions, to gain insights into the characteristics of the environmental source. This study employed broadband cavity-enhanced absorption spectroscopy to detect gaseous CHOCHO, methylglyoxal, and $NO_2$. Two different detection methods are compared. Spectrograph and CCD Detection: This approach involves coupling the system to a spectrograph with a charge-coupled device (CCD) detector. It achieved a 1 min 1-σ detection limit of $2.5 \times 10^8$ molecules/cm$^3$, or 10 parts per trillion (ppt). Methylglyoxal and $NO_2$ achieved 1 min 1-σ detection limits of 34 ppt and 22 ppt, respectively. Interferometer and PMT Detection: In this method, an interferometer is used in conjunction with a photomultiplier tube (PMT) detector. It resulted in a 2 min 1-σ detection limit of $1.5 \times 10^{10}$ molecules/cm$^3$, or 600 ppt. The $NO_2$ 2 min 1-σ detection limit was determined to be 900 ppt. Concentrations of methylglyoxal were difficult to determine using this method, as they appeared to be below the detection limit of the instrument. This study discusses the advantages and limitations of each of these detection methods.

## 1. Introduction

Glyoxal (CHOCHO), also known as ethandial or oxaldehyde, is a trace gas common in the atmosphere that can be used as an indicator of biogenic emissions [1,2]. CHOCHO is produced through the oxidation of various volatile organic compounds (VOCs) emitted by vegetation [2–4]. As a result, CHOCHO can be used as a robust indicator of biogenic VOC emissions. These emissions are crucial to monitor as they contribute to the overall VOC concentration in the atmosphere and cannot be controlled like anthropogenic VOC emissions. Anthropogenic VOC emissions result in the formation of formaldehyde ($CH_2O$) [5]. Therefore, $CH_2O$ and CHOCHO concentrations are often ratioed to understand the relative proportion of hydrocarbons that are emitted either from biogenic or anthropogenic sources. [5–12]. Recent studies indicate that anthropogenic emissions can also be a source of CHOCHO in large urban areas, known as megacities, particularly during the winter [13,14]. In these areas, CHOCHO is primarily produced by fossil fuel combustion from vehicles [15]. On-road CHOCHO emissions, however, have been found to be 38 times lower than $CH_2O$ emissions [16,17]. CHOCHO can also be emitted directly into the atmosphere through biomass burning [5,18,19]. Notably, CHOCHO emissions from biomass burning are approximately twice as high as $CH_2O$ emissions [20]. CHOCHO has a short atmospheric lifetime of a few hours and has been employed to identify photochemical hotspots on a global scale [21]. Monitoring VOC emissions is essential for understanding local atmospheric ozone formation pathways [22]. Furthermore, CHOCHO has been identified as a precursor for secondary organic aerosol (SOA) formation [23]. Several studies have employed various instruments to measure glyoxal (CHOCHO) and related compounds on a global scale. Wittrock et al. utilized a multi-axis differential optical absorption spectroscopy (MAX-DOAS) instrument and a scanning imaging absorption spectrometer for atmospheric cartography (SCIAMACHY), finding slant columns above a threshold of $5 \times 10^{14}$ molecules·cm$^{-2}$ to be approximately $4.2 \times 10^{15}$ molecules·cm$^{-2}$ [5]. In Australia, Ryan et al. achieved a detection limit of $8.84 \times 10^{13}$ molecules·cm$^{-2}$ using a MAX-DOAS instrument [6]. Vrekoussis et al. utilized the Global Ozone Monitoring Experiment-2 (GOME-2) instrument on the MetopA satellite, reporting detection limits on the order of $10^{13}$ molecules·cm$^{-2}$ [10]. Another study by Vrekoussis et al. employed differential optical absorption spectroscopy (DOAS) to compute a slant column density of $6.23 \times 10^{15}$ molecules·cm$^2$ [21]. Hong et al. measured CHOCHO over China with a MAX-DOAS instrument, reporting a detection limit of 20 ppt [11]. Chen et al. used the TROPOspheric Monitoring Instrument (TROPOMI) on a satellite, achieving a detection limit of $2 \times 10^{15}$ molecules·cm$^{-2}$ [8].

In addition to MAX-DOAS and satellite-based instruments, other techniques were employed for glyoxal measurement. Thalman et al. utilized light-emitting diode cavity-enhanced differential optical absorption spectroscopy (LED-CE-DOAS) with a 1 min normalized detection limit of 19 ppt [24]. DiGangi et al. measured CHOCHO using a Madison-Laser Induced Phosphorescence (Mad-LIP) instrument, reporting a 3σ detection limit of 16 ppt for a 1-s integration time [9]. Wang et al. employed 2,4-dinitrophenylhydrazine-high-performance liquid chromatography (DNPH-HPLC) for glyoxal measurement, reporting concentrations in the tens of ppt range [23]. Kaiser et al. used an airborne cavity-enhanced spectrometer (ACES) with 6% accuracy and 32 ppt precision for CHOCHO measurement [7]. Liu et al. utilized broadband cavity-enhanced absorption spectroscopy (BBCEAS) to measure glyoxal, reporting a 30 ppt detection limit and 10% accuracy for a 1 min integration time [12]. Similarly, Fang et al. and Washenfelder et al. employed BBCEAS, reporting 1σ detection limits of 28 ppt and 29 ppt, respectively, for a 1 min integration time [25,26].

BBCEAS operates as an intensity-dependent technique, involving the coupling of a continuous wave of broadband light into a high-finesse cavity constructed with two highly reflective mirrors [27–30]. The technique strategically leverages the Beer-Lambert law to enhance detection limits by extending the path length of light through the sample. Achieving multiple passes through the sample is facilitated by directing light into the cavity, where it undergoes numerous reflections before eventually exiting through the backside of the cavity mirrors. Subsequently, this emitted light is directed into a detector, measuring its relative intensity both before the introduction of an absorber into the cavity and in the presence of the absorber.

Mode matching refers to the alignment of the laser mode's resonance with the optical cavity, facilitating easy coupling [31]. To couple the lower-order modes of the light source, it is necessary to match the beam size and convergence with specific cavity parameters, such as the radius of curvature of the mirrors and cavity length. In BBCEAS, wavelength selection occurs after the cavity, eliminating the need for spatial mode matching [32]. When utilizing an incoherent broadband light source covering the desired range, the light inevitably contains eigenmodes for the cavity with specific parameters, while higher-order modes are lost in the cavity. This implies that although the coupling efficiency may be low or not entirely meet the requirements to be totally mode-matched, like in cavity ring-down spectroscopy, a fraction of the light from the incoherent source will always couple into the cavity. Therefore, even if the mirrors experience slight thermal fluctuations or minor vibrations and instabilities, there will still be modes within the incoming light that match the cavity. Lehmann and Romanini further demonstrated that the superposition principle is maintained within the cavity, ensuring that transmitted light does not exhibit mode structure when detected with moderate resolution [33].

In this study, an intercomparison is made between two instruments used to measure glyoxal concentrations. Broadband cavity-enhanced absorption spectroscopy (BBCEAS) was coupled with two different detection techniques: 1. Spectrograph with a charge-coupled device (CCD) detector 2. Interferometer with a photon multiplier tube (PMT) detector. This comparison aims to identify a path to a cost-effective and potentially less data-intensive method for measuring trace gases of interest in the atmosphere. The focus is on detecting glyoxal, a significant gas associated with biomass burning and biogenic emissions, as discussed earlier. This research represents a notable improvement in glyoxal detection, achieving lower detection limits for local ground-level concentrations compared to prior studies. It also describes a cheaper method of detection with similar detection limits to those in previous studies.

## 2. Materials and Methods

### 2.1. Experimental Setup

The full experimental setup is shown in Figure 1. The 92.5 cm BBCEAS cavity used for CHOCHO measurement consists of two 1″ highly reflective mirrors (Layertec) encased in a carbon rod cage, secured with 3D-printed brackets. The mirrors had a radius of curvature of 1000 mm, and the reflective coating was active between 430 and 480 nm and centered at 455 nm. An LED light source (Osram LZ1-10B202-0000, Mouser Electronics, Mansfield, TX, USA), centered at 450 nm, is temperature-controlled (TEC-1091, Meertstetter Engineering, Rubigen, Switzerland) using a Peltier cooler (CP30138, Digi-Key Electronics, Thief River Falls, MN, USA), maintaining $\pm 0.01$ °C precision. This Peltier cooler is attached to a copper heat sink to which a fan is attached (E97379, Intel, Santa Clara, CA, USA). Temperature-controlling the LED is essential to minimizing noise due to thermal fluctuations of the cavity's light source. Thermal fluctuations of the light source will cause changes in the emission intensity of the LED. This, if untracked, would lead to offsets relative to the reference spectrum or even an artificial structure in the extinction coefficient, depending on the LED. This could then lead to skewed concentrations or issues in spectral fitting, which would negatively affect the instrument's detection limit through the injection of noise into the data. The LED light is directed into the cavity through a 1-inch F1 collimating lens (Thorlabs, Newton, NJ, USA) and the back of the first mirror. The cavity is enclosed by a ¾″ Teflon tube (PFA McMaster-Carr, Elmhurst, IL, USA), affixed to the 3D-printed brackets at both ends. Calibrated mass flow controllers (202 Series, Teledyne, Thousand Oaks, CA, USA) regulate all gas flows. Flow controllers were calibrated using a Mesa Labs Defender 530+. CHOCHO is introduced into the cavity by bubbling $N_2$ through a 40% *w/w* aqueous CHOCHO solution (A16144-AP, Thermo Scientific, Waltham, MA, USA). $NO_2$ is introduced via a custom cylinder containing a 60.3 ppm standard (Air Liquide America Specialty Gases LLC, Houston, TX, USA), merged into the air flow entering the cavity. Light leaking through the cavity's second mirror is focused by a 1-inch F4 focusing lens through a bandpass filter (ThorLabs, Newton, NJ, USA), centered at 450 nm with a 40 nm full-width half-maximum (FWHM), into a bifurcated fiber optic (BF19Y2HS02, Thorlabs, Newton, NJ, USA). One end of the bifurcated fiber optic connects to a spectrograph (Shamrock 303i, Andor, Andor, Belfast, Northern Ireland, UK) equipped with a 100 µm slit, a 1200 lines/mm grating, and a cooled CCD (iDus, Andor) set at $-20$ °C for detection. The other end is coupled into a smaller cage housing a 1″ F4 collimating lens (ThorLabs), a bandpass interferometer (12-277, Edmund Optics, Barrington, NJ, USA) centered at 457.9 nm with a 1.7 nm FWHM, mounted on a pivoting optic mount (CP360Q, Thorlabs), and a ½″ F0.5 focusing lens (ThorLabs). The pivoting optic mount is controlled by a servo motor (MG90S 9g) driven by an Arduino Nano. The output from this cage is linked via a fiber optic (ThorLabs FG910LEC) to a PMT (P30A-14, Sens Tech, Egham, Surrey, UK) biased at 1150 V. The PMT signal is then directed to a photon counter (SR400, Stanford Instruments, Sunnyvale, CA, USA) for measurement, and the output voltage is recorded on a analog-to-digital converter (ADC) (U6, Labjack, Lakewood, CO, USA).

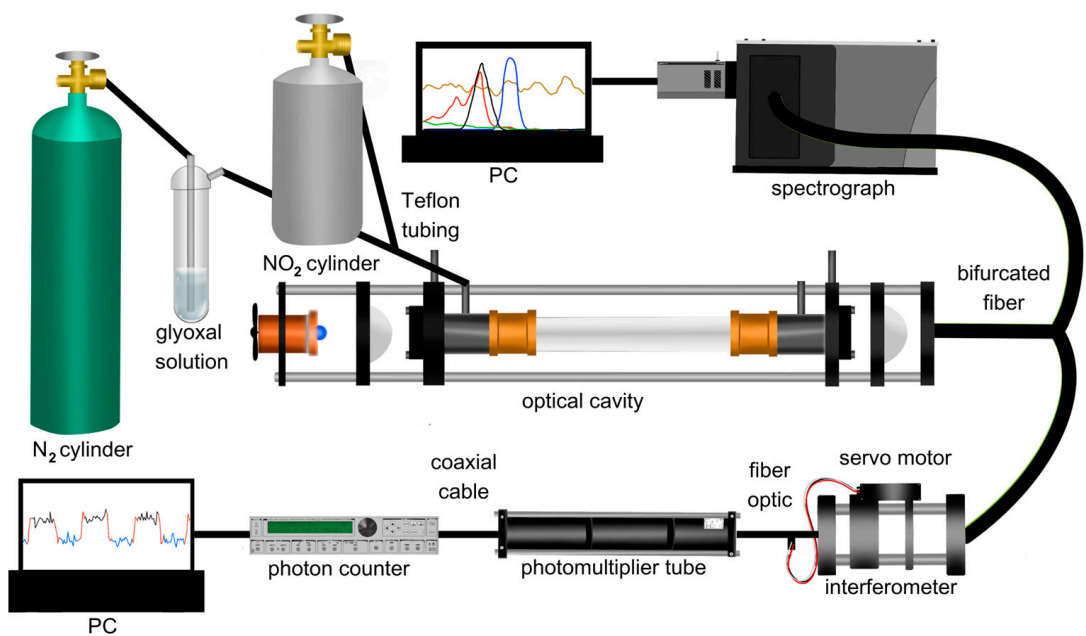

**Figure 1.** Experimental setup showing the BBCEAS with the spectrograph setup (above cavity) and interferometer setup (below cavity).

*2.2. Interferometer Setup*

Fabry–Perot etalons and other interferometers have been widely used for atmospheric gas sensing due to their high spectral resolution, improved light throughput compared to grating spectrographs, and field-ready compactness [34,35]. In this work, a bandpass filter is employed as an interferometer, functioning in a similar manner. Figure 2 illustrates the two windows created by the interferometer's turning for on-band (in black) and off-band (in blue) CHOCHO detection. The on-band, centered at 455.5 nm, corresponds to the spectral region overlapping with CHOCHOs absorption cross-section, while the off-band, centered at 458 nm, represents a region where CHOCHOs absorption is minimal. An inherent limitation of this approach is its inability to distinguish between multiple absorbers without a correction for their respective ratios. For instance, $NO_2$ absorbs across all the investigated wavelengths, affecting both on-band and off-band windows similarly. Conversely, CHOCHO exhibits a substantial difference in absorption between on-band and off-band windows. This large differential makes it easier to tease out the CHOCHO absorption on the on-band and define a CHOCHO concentration. Methylglyoxal's absorption is distinct from CHOCHOs and impacts the 453–458 nm transmission window more substantially than the 457–459 nm transmission window. Given methylglyoxal's much smaller absorption cross-section (15× smaller than CHOCHOs), corrections can be applied for the co-detection of both species. In a field experiment, narrower windows can be employed for specific absorption cross-sections, such as hydroxyl radical's, resulting in minimal interference from other absorbers. Scanning multiple windows instead of using only on-band and off-band windows enables low-resolution fitting of absorption cross-sections and species identification.

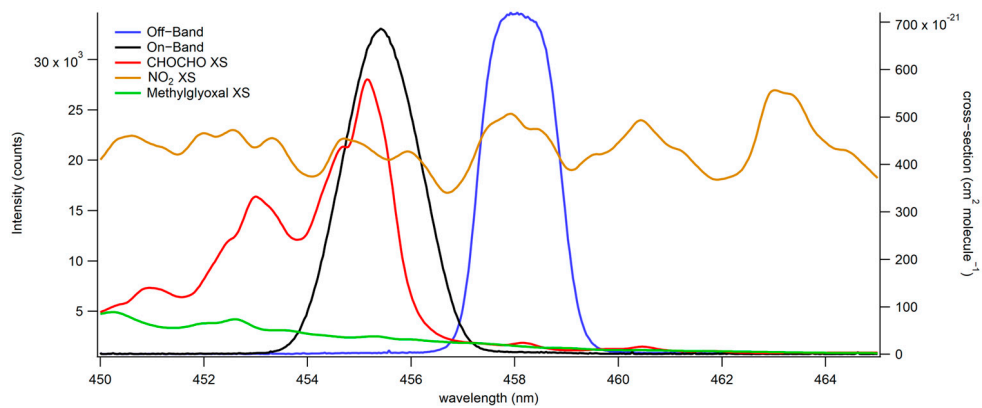

**Figure 2.** Absorbing species' absorption cross-sections indicate the absorption on-band and off-band. This is used for the on-band (in black) and off-band (in blue) detection of CHOCHO (absorption cross-section shown in red) and methylglyoxal (absorption cross-section shown in green). NO$_2$'s absorption cross-section is shown in brown.

*2.3. Data Processing*

2.3.1. Spectrometer Data

Data collected using the spectrometer were analyzed using the traditional BBCEAS formulae and spectral fitting of absorption cross-sections to determine the concentrations of absorbers present [36,37]. The reflectivity of the mirrors in the cavity was determined using the Rayleigh scattering of ultra-pure He and N$_2$. This is described in Equation (1), where d$_0$ is the cavity length, $\epsilon_{Ray}$ is the extinction due to Rayleigh scattering of the respective gases, and I is the spectral intensity while each gas is flowing [24,37,38]. Equation (2) describes the calculation of the extinction coefficients using the intensity of the spectra, where $\epsilon(\lambda)$ is the wavelength-resolved extinction, R($\lambda$) is the reflectivity, $\epsilon$ is the extinction due to Rayleigh scattering, I$_0$ is the reference spectrum, and I is the measurement spectrum. The extinction coefficients were then used to solve for the concentrations of known absorbers using non-linear least-squares fitting, as seen in Equation (3) [26].

$$R_{(\lambda)} = 1 - d_0 \frac{\left(\frac{I_{N_2(\lambda)}}{I_{He(\lambda)}}\right)\epsilon_{Ray\ (\lambda)}^{N_2} - \epsilon_{Ray\ (\lambda)}^{He}}{1 - \left(\frac{I_{N_2(\lambda)}}{I_{He(\lambda)}}\right)} \quad (1)$$

$$\epsilon_{(\lambda)} = \left(\frac{1 - R_{(\lambda)}}{d_0} + \epsilon_{Rayleigh,\ N_2\ (\lambda)}\right)\left(\frac{I_0\ (\lambda) - I_{(\lambda)}}{I_{(\lambda)}}\right) \quad (2)$$

$$\epsilon_{(\lambda)} = \sigma_A[A] + \sigma_B[B] + \sigma_C[C] + polynomial \quad (3)$$

2.3.2. Interferometer Data

In this experiment, only one PMT and one interferometer were used. The interferometer was programmed to rotate to pivot between the on-band (453.15–457.5 nm) and off-band (456.5–459.95 nm) windows. Due to the turning of the interferometer, the resultant data possesses what looks like a square wave function, as seen in Figure 3. This means that data for the higher intensity on-band signal (453.15–457.5 nm, in black) had to be separated from the lower intensity off-band signal (456.5–459.95 nm, in blue) and any signal that is acquired while the interferometer is changing positions (in red). The data were parsed accordingly, as shown in Figure 3, before calculating extinction coefficients.

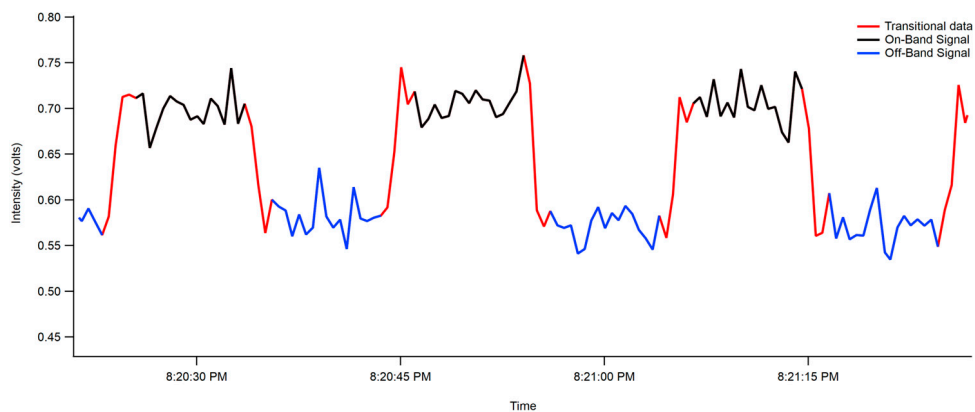

**Figure 3.** On- and off-band data of the interferometer as measured by the PMT. The separation of on-band signal (453.15–457.5 nm, in black) being separated from off-band signal (456.5–459.95 nm, in blue) and transitional signal acquired while the interferometer is changing positions (in red).

The interferometer data are processed similarly to the spectrograph data, with one key difference: instead of calculating an extinction coefficient for each individual wavelength, an extinction coefficient is calculated for the central wavelength of each window. In this case, we compute extinction coefficients for both 458 nm and 455.5 nm.

Next, we integrate the absorption cross-section for each absorber across each window, resulting in absorption values at each window position. These absorption values are then combined with the respective extinction coefficients to determine the concentrations of the absorbers.

While Equations (1) and (2) are applied in the same manner as described in the spectrograph data, with the only variation being the independent calculation of each wavelength (455 and 458), Equation (3) transforms into Equation (4) as follows:

$$\epsilon_{458\text{ nm}} = [CHOCHO]\sigma_{CHOCHO_{485\text{ nm}}} + [NO_2]\sigma_{NO_{2_{458\text{ nm}}}} \tag{4}$$

Since methylglyoxal and CHOCHO have no contribution at 458 nm because both of their absorption cross-sections go to zero, Equation (4) may be rewritten as Equation (5):

$$\epsilon_{458\text{ nm}} = [NO_2]\sigma_{NO_{2_{458\text{ nm}}}} \tag{5}$$

Equation (4) will also change at 455 nm to take the form of Equation (6), where the $NO_2$ concentration from the BBCEAS is substituted in and the CHOCHO concentrations are solved for:

$$\epsilon_{455\text{ nm}} = [CHOCHO]\sigma_{CHOCHO_{455\text{ nm}}} + [NO_2]\sigma_{NO_{2_{455\text{ nm}}}} \tag{6}$$

2.3.3. Error Analysis

The uncertainty associated with data collection using BBCEAS has previously been defined as 1.3% at the center of the mirrors and 2% towards the edges of the mirrors [38]. We have chosen to propagate 2% as a worst-case scenario. The uncertainty associated with the absorption cross-sections for glyoxal, methylglyoxal, and $NO_2$, respectively, is 5%, 5%, and 2%, as reported [39–42]. These absorption cross-section uncertainties were combined in quadrature with the error associated with the optical cavity to produce uncertainties for glyoxal, methylglyoxal, and $NO_2$ of 5.4%, 5.4%, and 2.8%, respectively. The effect of aerosols on these measurements by the spectrometer/CCD setup has also been reported in previous work [37]. For the interferometer setup, each absorber's relative absorbance cross-sections at various wavelengths were compared to sum up the total extinction coefficient at each wavelength. This approach provided the contribution of each absorber to the total extinction coefficient over time, allowing the extraction of each absorber's contribution.

This correction method can also be applied to broadband absorbers with minimal variation between the two wavelength bands, such as aerosols, and can help account for background noise issues like light source drift. This closed-cavity configuration of this instrument will always be accompanied by a particle filter upstream of the cavity. aerosol could be accounted for with assumptions about the relative contribution of $NO_2$ and aerosol. The wavelength dependence of aerosol extinction has been shown to be calculated as $\lambda^{-1.5}$, which can be used to calculate the extinction due to aerosol at each wavelength [24]. Formulae for the calculation of the extinction due to aerosol were also defined in the same work and could be used if an open-cavity version of this instrument were to be made. We can, therefore, calculate a simulated aerosol extinction for our on-band signal (455 nm) and off-band signal (458 nm) to be $1.029 \times 10^{-6}$ cm$^{-1}$ and $1.019 \times 10^{-6}$ cm$^{-1}$, respectively, which vary by <1% and would be applied as a direct intensity correction similar to $NO_2$ when retrieving the CHOCHO concentrations.

## 3. Results and Discussion

*3.1. Dual Detection (Spectrometer/CCD and Interferometer/PMT)*

3.1.1. Spectrometer/CCD

The mirror reflectivity, as defined in Equation (1), was determined to be 0.99985 at a wavelength of 455 nm. The shape of the reflectivity curve, illustrated in Figure 4, results from the reflectivity of the coated mirrors at different wavelengths within the mirrors' dielectric coating. This favorable signal-to-noise ratio was preserved during the calculation of the extinction coefficients, as outlined in Equation (2). This preservation led to a robust fit to the extinction coefficients, resulting in a low and unstructured fit residual, as depicted in Figure 5.

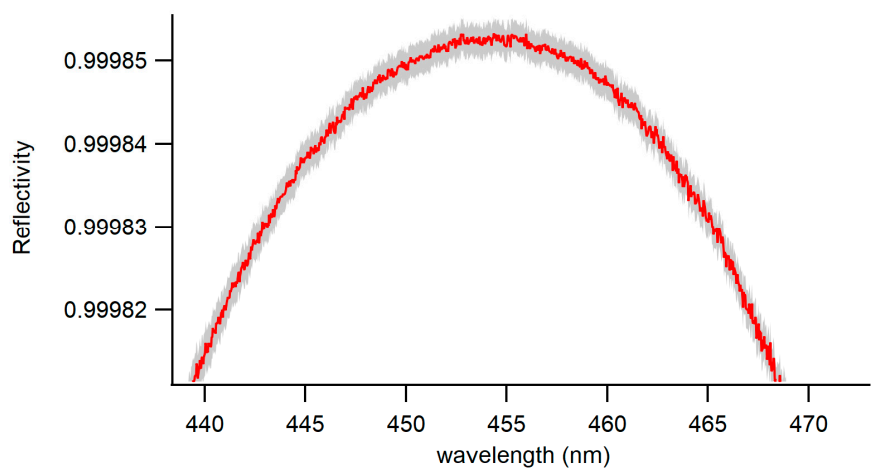

**Figure 4.** Reflectivity curve of the BBCEAS cavity as a function of wavelength (red). Uncertainty associated with the measured values is shown in gray.

Both CHOCHO and methyl-glyoxal were successfully fitted individually, as demonstrated in Figure 5A, and in the presence of other potent absorbers, as shown in Figure 5B. The 1 min 1-σ detection limit of the BBCEAS coupled to the spectrograph for CHOCHO is $2.5 \times 10^8$ molecules per cubic centimeter, or 10 parts per trillion (ppt). Methylglyoxal and $NO_2$ achieved 1 min 1-σ detection limits of 34 ppt and 22 ppt, respectively.

The successful fit resulted in consistently smooth concentration plots throughout the experiment. Initially, CHOCHO was introduced into the cavity at five concentrations, each in the order of $10^{10}$ molecules/cm$^3$, or single-unit parts per billion (ppb). Subsequently, CHOCHO flow remained constant while $NO_2$ was introduced into the cavity, as illustrated in Figure 6A.

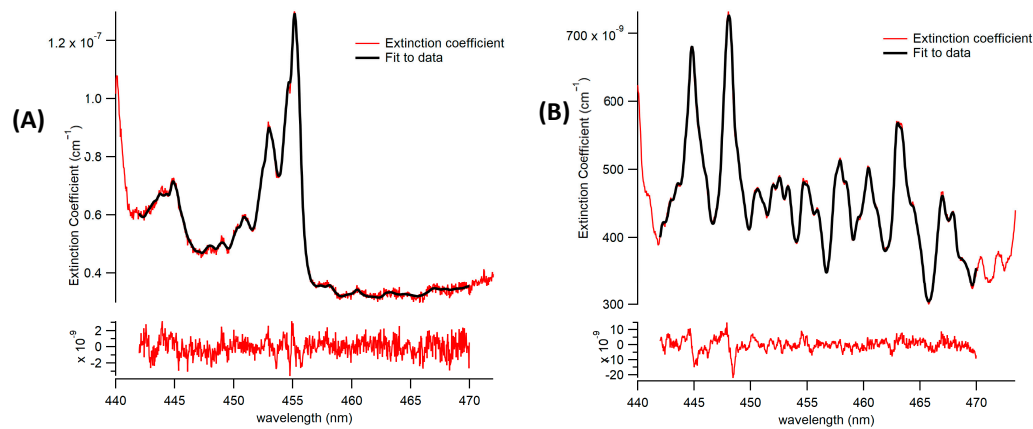

**Figure 5.** Fit (in black) to measured extinction (in red) and fit residuals using the spectrometer. (**A**) A representative fit of CHOCHO and methylglyoxal is being detected. (**B**) A representative fit of co-detection of CHOCHO and methylglyoxal in the presence of $NO_2$.

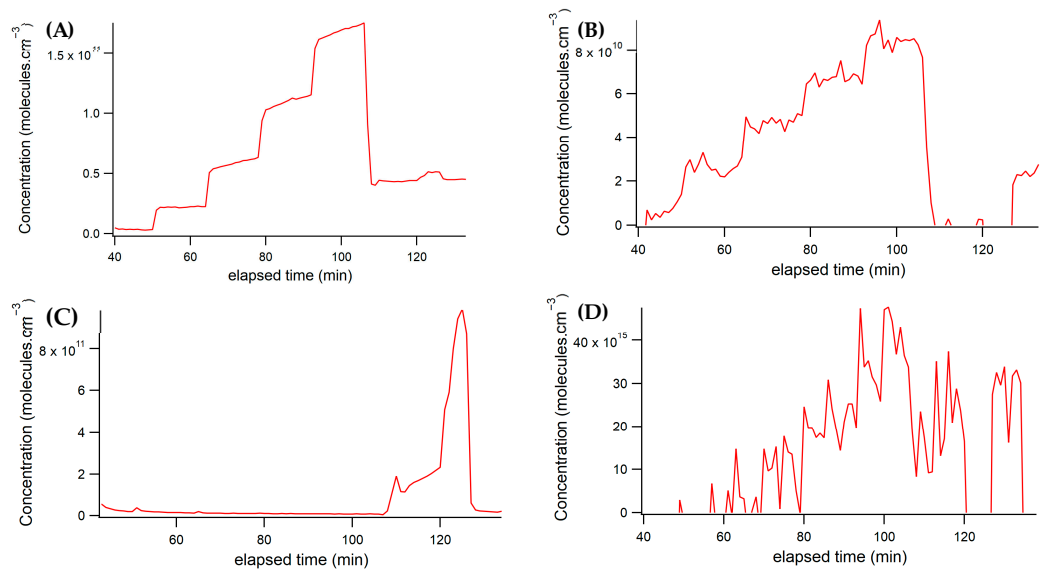

**Figure 6.** Concentration vs. spectrum number plots show data acquired from the spectrometer/CCD set-up. (**A**) CHOCHO concentration from the spectrograph; (**B**) Methylglyoxal concentration from the spectrograph; (**C**) $NO_2$ concentration from the spectrograph. (**D**) Water vapor concentration from the spectrograph.

Notably, due to sample contamination with methylglyoxal, methylglyoxal concentrations fluctuated in response to changes in CHOCHO concentrations. Although these variations are not as clearly defined, they are visible in Figure 6B.

Throughout most of the experiment, $NO_2$ was intentionally kept off. Toward the end, two different concentrations of $NO_2$ were introduced into the cavity, while CHOCHO flow remained stable. The flow controller for the $NO_2$ standard experienced some time-dependent variability, resulting in a range of concentrations on the order of $10^{11}$ molecules/cm$^3$ or tens to hundreds of ppb of $NO_2$ in the sample. The $NO_2$ concentrations throughout the experiment are depicted in Figure 6C.

Additionally, water vapor was measured and is shown in Figure 6D. Its temporal variation followed a pattern like glyoxal, depending on the volume of air passed through the bubbler.

### 3.1.2. Interferometer/PMT

Figure 7 displays the unprocessed CHOCHO concentration data obtained from the PMT. The act of reading out the voltage from the photon counter onto the ADC introduced some noise into the PMT data. Moreover, the experimental setup for the interferometer added its own noise. Consequently, the 2 min 1-$\sigma$ detection limit for CHOCHO based on the interferometer data are $1.5 \times 10^{10}$ molecules/cm$^3$, or 600 parts per trillion (ppt). The NO$_2$ 2 min 1-$\sigma$ detection limit was determined to be 900 ppt. Concentrations of methylglyoxal were difficult to determine using this method, as they appeared to be below the detection limit of the instrument. While this detection limit remains favorable for measuring ambient CHOCHO concentrations, it is essential to note that the setup introduced noise sources that could potentially be mitigated if access to alternative equipment were possible. Examples may be a better PMT or a PMT that provides a voltage output and not a photon-counting PMT, better lenses that have better F-matching, and a stronger light source that could provide more initial photons. All of these would allow for more signal and less noise.

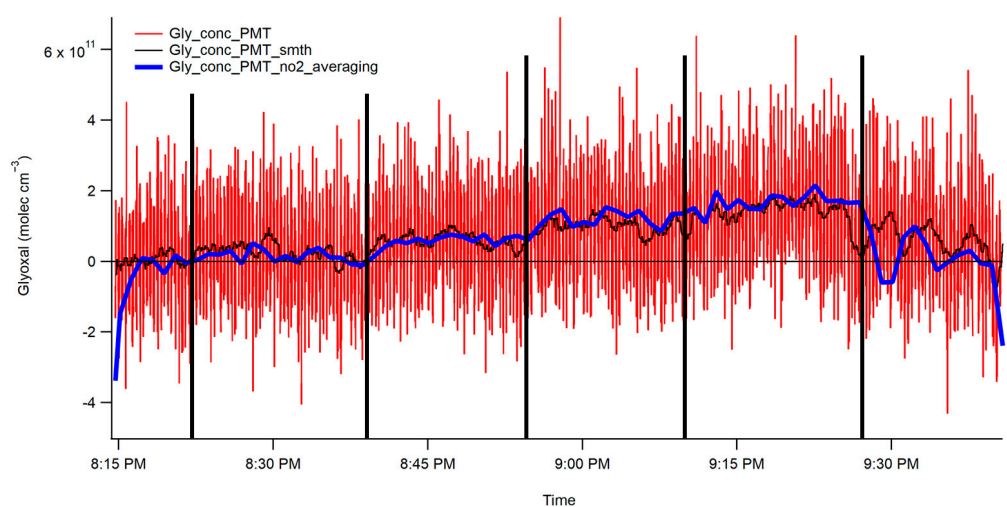

**Figure 7.** CHOCHO fits the interferometer/CCD data. The five vertical black lines indicate the end of each of the five steps in the concentration of CHOCHO. The blue trace represents the two-minute averaged data, and the black trace represents the rolling boxcar averaged data. The vertical black lines are to draw attention to concentration changes.

The plot showcases the five CHOCHO concentrations and highlights the interference from NO$_2$ in the final section. The CHOCHO concentration was determined using two methods: first, a rolling boxcar method, displayed in black, was employed to smooth the raw data. Second, the data were averaged over two minutes and then plotted. These two smoothing techniques were subsequently compared to the spectrograph concentrations, as shown in Figure 8. The NO$_2$ trace of two-minute averaged data closely follows the trends in the spectrograph's data. It is worth noting that the CHOCHO data exhibits more noise compared to the spectrograph data; however, both the two-minute averaged data and the NO$_2$ averaged data effectively capture the trends seen in the spectrograph data.

As previously discussed in the Spectrometer section, the analysis identified the presence of methylglyoxal and water in the sample emanating from the CHOCHO solution in the bubbler. In the context of interferometer measurements, the interference posed by methylglyoxal is minimal. This is because the absorption cross-section of methylglyoxal is approximately 15 times smaller than that of CHOCHO at the on-band position.

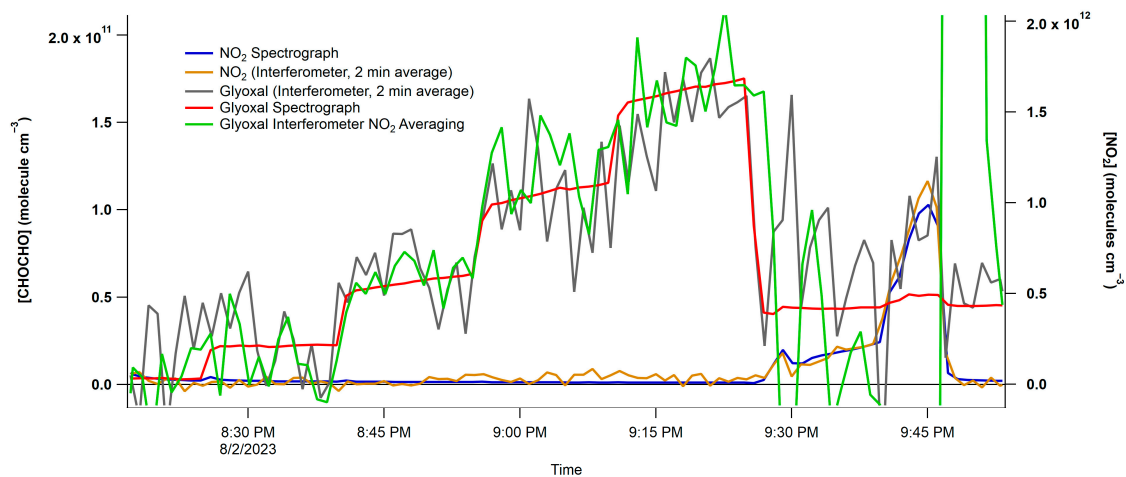

**Figure 8.** Concentration vs. spectrum number plot from the interferometer/PMT and spectrograph/CCD. The blue trace is the spectrograph/CCD $NO_2$ data, and the purple trace are the interferometer/PMT $NO_2$ data.

Conversely, water vapor has distinct absorption bands between 440 and 450 nm and from 466 to 476 nm. Both of the interferometer's bandpass positions fall within these two absorption windows. Considering the concentrations of methylglyoxal determined with the spectrograph/CCD, it can be estimated that roughly 1/30 of the CHOCHO measured in the interferometer/PMT set-up could be attributed to methylglyoxal.

Due to the interferometer's ability to measure primarily at two wavelengths, distinguishing between more than two species that absorb in this region is not possible. However, this is also seen in other cavity instruments such as cavity attenuated phase shift (CAPS) spectroscopy and cavity ringdown spectroscopy (CRDS) that measure at a single wavelength. Measuring at two wavelengths allows the cancellation of broadband absorbers if the extinctions are similar at both wavelengths. This has been conducted in previous studies to retrieve aerosol concentrations using a similar instrument [43].

## 4. Conclusions

In this study, a ¾″ BBCEAS cavity equipped with 1″ mirrors was employed for the simultaneous detection of CHOCHO, methylglyoxal, and $NO_2$ using two distinct detection methods. The first method involved coupling the BBCEAS to a spectrograph with a CCD detector. This method yielded successful results, with a strong fit to the measured extinction and a small fit residual. Consequently, it achieved a 1 min 1-σ detection limit of $2.5 \times 10^8$ molecules per cubic centimeter, equivalent to 10 parts per trillion (ppt). Methylglyoxal and $NO_2$ achieved 1 min 1-σ detection limits of 34 ppt and 22 ppt, respectively.

The second method entailed the use of an interferometer coupled to a PMT to measure CHOCHO. This approach exploited the difference in absorption between two distinct windows generated by the interferometer. The first window allowed "on-band" light (453.15–457.5 nm), which CHOCHO absorbed, while the second window permitted "off-band" light (456.5–459.95 nm), which CHOCHO did not absorb. Through this second method, a 2 min 1-σ detection limit of $1.5 \times 10^{10}$ molecules per cubic centimeter, or 600 ppt, was achieved. The $NO_2$ 2 min 1-σ detection limit was determined to be 900 ppt. Concentrations of methylglyoxal were difficult to determine using this method, as they appeared to be below the detection limit of the instrument. It is worth noting that while this setup detected an interference with methylglyoxal, methylglyoxal is typically not a primary atmospheric measurement due to its poorer detection limits and faster photolysis rates, leading to lower concentrations, except in specific cases like wildfire plumes.

Although the detection limit achieved through the interferometer method is favorable and can be used for ambient species detection, it may be relatively lower than the spectrograph method, partially due to equipment limitations. Interferometer limitations include lower time resolution compared to the spectrograph and challenges in separating multiple absorbers that may absorb within the same wavelength window. The first limitation could potentially be mitigated by using a second PMT setup, allowing simultaneous detection of on-band and off-band signals. However, this would be constrained by the intensity of the light source and its ability to produce sufficient photons for both PMTs.

To address the second limitation, each absorber's relative absorbance cross-sections at various wavelengths were compared to sum up the total extinction coefficient at each wavelength. This approach provided the contribution of each absorber to the total extinction coefficient over time, allowing the extraction of each absorber's contribution. This correction method can also be applied to broadband absorbers with minimal variation between the two wavelength bands, such as aerosols, and can help account for background noise issues like light source drift.

When targeting a specific species, the wavelength bands can be positioned so that the species of interest is the only one absorbing in the on-band window. In cases where this is not feasible, the species of interest should have the most significant difference in absorbance between the on-band and off-band windows, with the smallest difference in wavelength. In situations where neither of these conditions can be met, as in this study, the absorbance cross-sections of the absorbers can be explicitly determined, assuming the presence of a second absorber (e.g., $NO_2$), or ratioed to account for an extinction process that is consistent between the two band positions.

This work provides the proof-of-concept coupling of etalon-based on/off band detectors with high-finesse optical cavities. This technique can provide detection for compounds with sharp drops in absorption cross-sections similar to CHOHCO, such as CHOH [44]. This instrument also holds the potential to serve as a path to a more cost-effective standard alternative for measuring local atmospheric concentrations of CHOCHO.

**Author Contributions:** C.E.F.: Writing of the original manuscript, experimental setup, data acquisition, data analysis. R.T.: Writing review and editing, aided in data analysis, experimental setup, conceptualization, and funding acquisition. M.C.A.: Writing review and editing, conceptualization, aided in experimental setup, and funding acquisition. J.C.H.: Conceptualization, writing review and editing, supervision, and funding acquisition. All authors have read and agreed to the published version of the manuscript.

**Funding:** This work was supported by the National Science Foundation, grant # 2114655.

**Institutional Review Board Statement:** Not applicable.

**Informed Consent Statement:** Not applicable.

**Data Availability Statement:** Available on BYU Scholar's archive https://scholarsarchive.byu.edu/data/56 (accessed on 28 August 2023).

**Conflicts of Interest:** The authors declare no conflicts of interest.